\documentclass[prb,twocolumn,showpacs]{revtex4}
\usepackage{epsfig,color}

\begin{document}
\title{Coupling of bonding and antibonding electron orbitals in double quantum dots by  spin-orbit interaction}
\author{M.P. Nowak } \affiliation{Faculty of  Physics and Applied
Computer Science, \\ AGH University of Science and Technology,\\ al.
Mickiewicza 30, 30-059 Krak\'ow, Poland}
\author{B. Szafran} \affiliation{Faculty of Physics and Applied
Computer Science, \\ AGH University of Science and Technology,\\ al.
Mickiewicza 30, 30-059 Krak\'ow, Poland}

\date{\today}

\begin{abstract}
We perform a systematic exact diagonalization study of spin-orbit coupling effects for stationary few-electron states confined in quasi two-dimensional double quantum dots.
We describe the spin-orbit-interaction induced coupling between bonding and antibonding orbitals and its consequences for magneto-optical absorption spectrum.
The spin-orbit coupling for odd electron numbers (one, three) %only weakly perturbs the ground-state wave functions.
%Nevertheless,
%the spin-orbit interaction
opens avoided crossings between low energy excited levels
of opposite spin orientation and opposite spatial parity. For two-electrons the spin-orbit coupling  allows for low-energy optical transitions that are otherwise forbidden by spin and parity selection rules. We demonstrate that the energies of optical
transitions can be significantly increased by an in-plane electric field but only for odd electron numbers.
Occupation of single-electron orbitals and effects of spin-orbit coupling on electron distribution between the dots are also discussed.
\end{abstract}
 \maketitle

\section{Introduction}
In a pair of quantum dots \cite{stacks,burkard,hu,harju}  defined in  semiconducting medium the charge carriers form extended wave functions when their tunneling through the interdot barrier becomes effective enough.
In vertically stacked quantum dots the extended electron and hole orbitals are probed by photoluminescence experiments in external electric field.\cite{stacks}
The electron single-dot orbitals hybridize to bonding ground states similar to the ones found in natural covalent molecules.
Recent studies\cite{climenteprb} indicated that the hole in artificial molecules of self-assembled quantum dots behaves in a different manner forming an antibonding
ground-state orbital. This peculiar behavior results\cite{climenteprb} of the spin-orbit (SO) coupling induced mixing  of light and heavy hole states.

In the present paper we study the mixing of bonding and antibonding electron orbitals that is induced by SO interaction
in planar systems of laterally coupled quantum dots.
The coupling between spatial and spin electron degrees of freedom results from  inversion asymmetry of the structure\cite{rashba} and/or the crystal lattice.\cite{dress} This asymmetry  enters into the two-dimensional
SO Hamiltonian which does not conserve the spatial parity
and couples the electron spin-up bonding orbitals with spin-down
antibonding orbitals.
In order to indicate experimentally accessible consequences of this coupling we consider
optical absorption spectra in the external magnetic field for up to three confined electrons.
In parabolic quantum dots the spin-orbit
coupling introduces a distinct dependence of the far infrared magneto-optical absorption spectra
on the number of confined electrons.\cite{czakra} We find that the SO
induced modification to the absorption spectra of double dots are qualitatively different for
even and odd electron numbers.

Laterally coupled quantum dots\cite{elzerman,petta} are considered candidates for realization of a quantum gate working on electron spins\cite{burkard} since
the height / width of the interdot barrier can be tuned by external voltages which is essential for the control of the spin exchange between the electrons confined in adjacent dots. The idea of the spin exchange motivated a number of theoretical investigations on the properties of electron systems in laterally coupled quantum dots.\cite{c1,c2,szafran,c4,c5,stopa,lb,pstano,baruffa,hindus}

The SO interaction is one of the issues that are investigated in the context of spin-based quantum information processing.\cite{esr1,esr2,esr3,kavokin,pstano,golo,baruffa,hindus,fw,fw2,us,us2,os1,os2,bednarek,foldi}
The SO coupling allows for spin manipulation by the spatial electron motion.\cite{os1,os2,bednarek,foldi}
Moreover, it leads to the spin relaxation\cite{esr1,esr2,esr3,pstano,golo} mediated by phonons, leading to information decay and decoherence.
Singlet-triplet induced avoided crossing of two-electron energy levels were observed in electron-transport spectroscopies for gated InAs nanowire quantum dots\cite{stxp}
as well as for double InAs quantum dots.\cite{stxp2}
The exchange interaction between electrons confined in separate dots  was found to contain an anisotropic component originating from the SO coupling,\cite{kavokin} which initially motivated a quest for spin processing procedures\cite{fw,fw2} minimizing its effects.
Later on, proposals of using the asymmetry of the exchange interaction for construction of universal quantum gates
that could work without single spin operations\cite{us,us2}  were formulated. Recently, it was demonstrated by the exact diagonalization that
\cite{baruffa}  the anisotropy of the exchange interaction -- previously discussed within approximate approaches -- is in fact absent in zero magnetic field.
%The smallness of the SO interaction energy turns out occasionally to be a misleading motivation for using a limited basis in calculation of confined states.
%Besides the problem of the exchange which at zero magnetic field seems anisotropic only in approximate treatment\cite{baruffa}
%for circular quantum rings the basis limited to the lowest radial state produces artifactally anisotropic charge density.\cite{nowak}
%In quantum dots the neglect of the excited Landau levels produces qualitatively wrong description of SO coupling effects at high magnetic field.\cite{szafran}
Therefore, the treatment of  spin-orbit coupling effects for double dots requires an exact diagonalization approach which we employ below.
%In the present work we apply the exact diagonalization to description of the spin-orbit coupling effects.
The SO coupled double quantum dots were so far studied by the exact diagonalization in Ref. [\onlinecite{ps2}], which provides
a detailed analysis of single-electron states and in Refs. [\onlinecite{baruffa,dest}] which deal with the electron pair in the
context of the exchange interaction.

%The control of the electron spin in GaAs based structures is to an extent hampered by the
%hyperfine interactions with spins of nuclei which produce a random fluctuating effective magnetic field of the order of mT. [petta]. The hyperfine field of the order of %mT is is negligible for the interdot tunnel coupling that is the topic of this work.

% To the best of our knowledge no description of the SO-induced mixing of bonding and antibonding electron orbitals was given so far.

%To the best of our no results for three-electrons, no results for optical transitions,
%- transitions with phonons in the context of SO induced spin-relaxation [esr].

\section{theory}

We consider an effective mass single-electron Hamiltonian of the form:
\begin{eqnarray}
h&=&\left( \frac{\mathbf{p}^2}{2m^*} + W({\bf r}) \right)\mathbf{1}\nonumber \\& +& \frac{1}{2}g\mu_BB\sigma_z + H_{SIA} + H_{BIA},
\end{eqnarray}
where $\mathbf{p}=\hbar\mathbf{k}=-i\hbar\nabla + e \mathbf{A}$,  $\mathbf{1}$ is the identity matrix, $W({\bf r})$ stands for the potential, $H_{SIA}$ and $H_{BIA}$ introduce Rashba\cite{rashba} (structure inversion asymmetry) and Dresselhaus\cite{dress} (bulk inversion asymmetry) spin-orbit interactions. The vector potential is taken in the symmetric gauge $A=\frac{B}{2}(-y,x,0)$.
The Rashba and Dresselhaus SO interactions have the form
\begin{equation}
H_{SIA}=\alpha \nabla W  \cdot (\sigma \times \mathbf{k}),
\end{equation}
%W=V(r) + |e|\mathbf{F}\cdot r\right
and
\begin{eqnarray}
  H_{BIA} &=& \gamma \left[ \sigma_{x} k_{x} (k^2_{z} - k^2_{y}) + \sigma_{y}k_{y}(k^2_{x} - k^2_{z})\right. \nonumber \\  &+& \left. \sigma_{z}k_{z}(k^2_{y} - k^2_{x})\right],
  \end{eqnarray}
  respectively. In Eqs. (2) and (3)
$\alpha$ and $\gamma$ are bulk SO coupling constants, $\sigma$'s are Pauli matrices and
 $x$, $y$, $z$ axes are oriented parallel to [100], [010] and [001] (growth) crystal directions, respectively.

We assume that the confinement potential forming the quantum dot is separable into an in-plane $V_c(x,y)$ and a growth direction $V_z(z)$ components
so that the potential appearing in the Hamiltonian (1) is
\begin{equation} W({\bf r})= V_c(x,y)+V_z(z) + |e|\mathbf{F}\cdot {\bf r} ,\end{equation}
where $\mathbf{F}$ is the electric field vector (below we always take $F_y=0$). In the following we adopt a two-dimensional approximation assuming that the electrons occupy a frozen lowest-energy state of quantization in the growth direction. The two-dimensional SO terms are obtained by averaging $H_{SIA}$ and $H_{BIA}$
over the wave function describing the electron localization in the growth direction. The two-dimensional Rashba terms  are usually \cite{dest} separated into a diagonal
\begin{equation}
H_{SIA}^{diag}=\alpha \sigma_z   \left( \left[ \frac{\partial V}{\partial y}  \right] k_x - \left[ \frac{\partial V }{\partial x} + |e|F_x \right] k_y \right),
\end{equation}
and
linear
\begin{equation}
H_{SIA}^{lin} = \alpha\langle\frac{\partial W}{\partial z}\rangle(\sigma_x k_y - \sigma_y k_x),
\end{equation}
parts.  In this formula the average gradient of the potential calculated for the wave function in the growth direction can be attributed to an effective $z$ component of the electric field
$F_z=\frac{1}{|e|}\langle\frac{\partial W}{\partial z}\rangle$.
The two-dimensional Dresselhaus SO interaction contains the linear
    \begin{equation}
H_{BIA}^{lin}=\gamma \langle k^2_z\rangle\left[\sigma_x k_x - \sigma_y k_y\right],
\end{equation}
and the cubic
 \begin{equation}
  H_{BIA}^{cub} = \gamma \left[\sigma_y k_y k^2_x - \sigma_x k_x k^2_y\right]
    \end{equation}
terms.
We assume that the quantum dot is made of In$_{0.5}$Ga$_{0.5}$As alloy for which we adopt the SO coupling constants
 $\alpha=0.572$ nm$^2$ (after Ref. \onlinecite{silva}) and $\gamma=32.2$ meVnm$^{3}$ (after Ref. \onlinecite{saikin}).
 The other material parameters are taken as arithmetic average\cite{will} of InAs and GaAs, i.e. we use the electron effective mass $m^*=0.0465m_0$, Land\'{e} factor $g=-8.97$ and dielectric constant $\epsilon=13.55$. The considered large value of the $g$ factor is in the order of the one found for in experimental
 samples\cite{stxp,stxp2} in which the SO coupling effects were studied. 
 
For the electron wave function in the growth direction identified with the ground-state of an infinite
rectangular potential well of height $d$ one obtains the two-dimensional linear Dresselhaus constant $\gamma^{2D}=\gamma\langle k^2_z \rangle =\gamma\frac{\pi^2}{d^2}$ [see Eq. (7)]. In the bulk of our calculations we assume a minimal but still realistic value of $d=5.42$ nm, for which $\gamma^{2D}=10.8$ meVnm.
%The cubic Dresselhaus interaction (7) as well as the diagonal Rashba term (5) introduce very small contributions to the results presented below.
%We found that
%the linear Rashba term does not qualitatively influence the present results (see Results Section), so unless stated otherwise for simplicity we assume %$F_z=0$.

%For consideration of case of equal strength of linear Rashba and Dresselhaus term we take $F_z=F_{z0}=188.8 \frac{keV}{cm}$.

\begin{figure}[ht!]
\epsfysize=50mm
                \epsfbox[4 213 600 570] {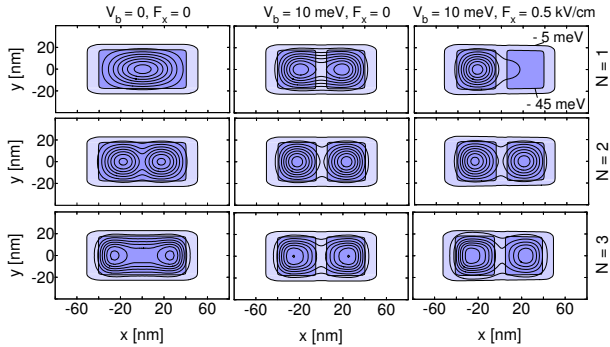}
                 \caption{The shades of blue show the in-plane potential of a single dot ($V_b=0$ -- left column of plots)
                 and of a double dot ($V_b=10$ meV -- central and right columns).
                                    In the right column of plots an in-plane electric field of $F_x=0.5$ kV/cm is included.
                                    Inside the light (darker) blue area the potential falls below -5 meV  (-45 meV).
                                    The contours indicate the charge density for a single (top row), two electrons (middle row) and three electrons (lowest row of plots) for $B=0$.}
 \label{confpot}
\end{figure}

The in-plane confinement potential is taken in form
\begin{eqnarray}
V_c(x,y)& =& -\frac{V_0}{\left( 1+ \left[ \frac{x^2}{R_x^2} \right]^\mu \right) \left( 1+ \left[ \frac{y^2}{R_y^2} \right]^\mu \right)}  \nonumber \\ &+& \frac{V_{b}}{\left( 1+ \left[ \frac{x^2}{R_b^2} \right]^\mu \right) \left( 1+ \left[ \frac{y^2}{R_y^2} \right]^\mu \right)},
\end{eqnarray}
where $V_0=50$ meV is the depth of the dots and $V_{b}$ is the height of the interdot barrier. We assume $\mu=10$ for which the potential profile
has a form of a nearly rectangular potential well, where
$2R_x=90$ nm and $2R_y=40$ nm determine the  size of the double dot in $x$ and $y$ directions respectively and $2 R_b=10$ nm is the thickness of the interdot barrier. We consider two values of the barrier height $V_b=10$ meV -- for the double dot potential and $V_b=0$ -- for a single
elongated dot. The elongated dot potential on the one hand corresponds to the limit case of strong interdot tunnel coupling and on the other
it is close in geometry to the nanowire quantum dots, in which the spin-orbit coupling induced singlet-triplet avoided crossing was observed
in a single-electron charging experiment.\cite{stxp}
The potential $V_c$ is displayed in Fig. \ref{confpot} for both the single and double dot.

The single-electron eigenfunctions are found by diagonalization of the two-dimensional version of Hamiltonian (1) in a basis of multicenter Gaussian functions\cite{chwiej} with embedded gauge invariance
\begin{equation}
\psi_{n} = \sum_{ks} c_{ks}^{n} \chi_{s}\exp \left[ -\frac{(\mathbf{r}-\mathbf{R_k})^2}{2a^2} + \frac{ieB}{2\hbar}(xY_k-yX_k)\right],
\end{equation}
where summation over $k$ runs over centers of Gaussian $\mathbf{R_k}=(X_k,Y_k)$,
  $s=\uparrow,\downarrow$ and $\chi_s$ are eigenstates of Pauli $\sigma_z$ matrix. The centers $\mathbf{R_k}$ are distributed on a rectangular mesh of $25 \times 11$ points spaced by $\Delta x = \Delta y = 5.2$ nm. The variationally optimal basis function parameter $a=4.7$ nm is used in the calculations.

The eigenproblem of of $N$-electron Hamiltonian \begin{equation} H=\sum_i^N h_i + \sum_{i=1,j>i}^N\frac{e^2}{4 \pi \epsilon_0 \epsilon r_{ij}}\end{equation} is solved using the configuration-interaction approach with a basis constructed of Slater determinants
built of single-electron eigenfunctions (10) of SO-coupled Hamiltonian. Convergence of the energies with a precision better than $1\mu$eV is usually reached for inclusion of thirty one-electron eigenstates.

The confinement potential (9) is symmetric with respect to the origin. In the present work the asymmetry effects are introduced by the in-plane electric field $F_x$. For $F_x=0$ and without SO coupling the stationary states possess a definite spatial parity with respect to point inversion $P\psi^n(-{\bf r})=\pm \psi^n({\bf r})$, where $P$ is the inversion operator. The eigenvalue $+1$ corresponds to even parity states and the eigenvalue $-1$ to the odd parity states. When SO is introduced the spatial parity eigenvalue is no longer a good quantum number even for $F_x=0$. For symmetric systems the SO coupled Hamiltonians commute with the operator $P\sigma_z$, which implies that the spin-up and spin-down components still possess definite but opposite spatial parities. We refer to $P\sigma_z$ as the s-parity operator. Eigenstates of this operator with eigenvalue +1 (-1)
are referred to as even (odd) s-parity  states or for brevity s-even (s-odd) states. The even s-parity states
have even-parity spin-up component and odd-parity spin-down component.

We evaluate the optical absorption spectrum using the energies of stationary states
 %with an assumption that the exciting light is circularly polarized.
and transition probabilities from state $k$ to $l$
that is proportional to the square of the dipole matrix element
\begin{equation}
I_{kl} = \langle \Psi_k | \sum_{j=1}^N \left( x_j \pm i y_j \right) | \Psi_l \rangle,
\end{equation}
where $\Psi_k$ is the $N$-electron wave function for $k$-th Hamiltonian (11) eigenstate and the signs $\pm$
correspond to opposite circular polarization of the exciting light.
%We investigate absorption probability of electromagnetic wave here the system undergo transition from the initial N-particle state $| \Psi_l \rangle$ %to the final $|\Psi_l\rangle$ state. The polarization of electromagnetic wave is plus(minus) for plus(minus) sign in equation(.)
The optical transitions conserve the electron spin and invert the spatial parity when it is a well-defined quantity.
When the SO coupling is introduced the optical transitions can only occur between states of opposite s-parity.

%For the s-parity operator eigenstates with the eigenvalue -1
%we refer to as odd s-parity or s-odd states.
\begin{figure}[ht!]
 \hbox{\rotatebox{0}{\epsfxsize=60mm
                \epsfbox[75 10 509 823] {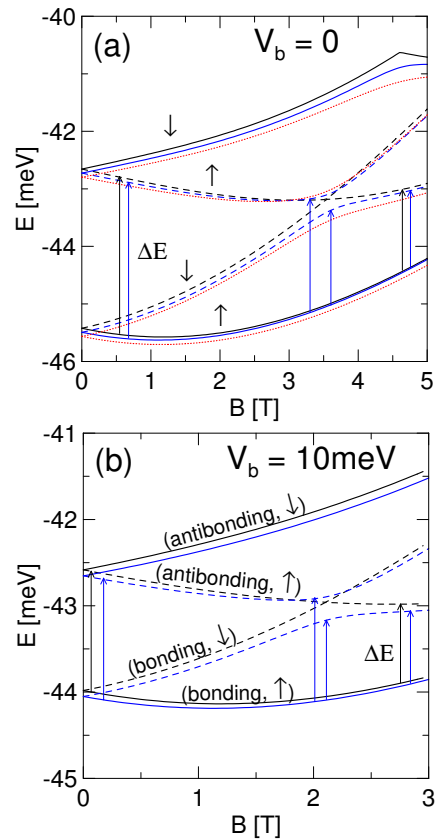}
                \hfill}}
\caption{Lowest single-electron energy-levels in function of the magnetic field
for a single elongated dot ($V_b=0$) (a) and for a double dot ($V_b=10$ meV) (b).
 Solid (dashed) lines correspond to the even (odd) s-parity.
  Black  lines show the results without  SO coupling. The blue curves show the results obtained with SO coupling
    for $F_z=0$ and the red curves in (a) for $F_z=188$ kV/cm.
  %The blue lines correspond to the set of parameters applied throughout the present paper for the SO coupling without linear Rashba term. The red lines in (a) correspond to a linear Rashba term as strong as the Dresselhaus one.
  The spin direction for the energy levels without SO coupling are marked with arrows.
The thin vertical lines indicate allowed optical transitions from the ground-state.
}
\label{bab1}
\end{figure}

\begin{figure}[ht!]
 \hbox{\rotatebox{0}{\epsfxsize=80mm
                \epsfbox[15 180 580 655] {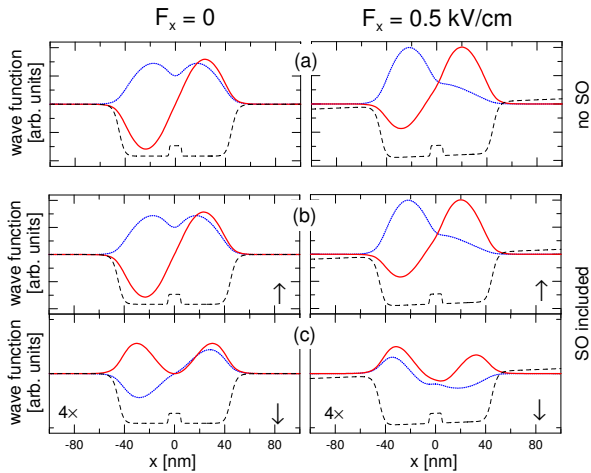}
                \hfill}}
\caption{
Dashed curves show the potential confinement profile for $V_b=10$ meV calculated for $y=0$.
(a) The spin-up components of the even s-parity ground state (red lines) and s-odd parity excited state (blue lines) in the absence
of SO coupling (spin-down components  exactly vanish). (b) Same as (a) but with SO coupling present. Spin-down components are presented
in (c) The scale for the wave function is the same on all the plots, but in (c) the wave functions were multiplied by 4. At right (left) panels an
electric field is $F_x=0.5$ kV/cm (zero). %We fixed a phase of the spinor wave function in a way that the spin-up component is purely real and the
%spin-down component is purely imaginary.
}
\label{bab}
\end{figure}

%Potencjal: single dot or double dot. Single - limit case of extreme interdot coupling.

%:to przerzucic
%Calculated absorption energies are of order of 1 meV which corresponds to the electromagnetic wave of nm length ($10^2$ GHz frequencies). Experiments %on absorption of such waves in nanostructures have been done, ie. [\cite{microfal}].

\section{results}

\subsection{Single electron}

The single-electron spectrum for a single elongated dot and for the double dot is presented in Fig. \ref{bab1}.
For $B=0$ the ground state and the first excited state are Kramers doublets. In each doublet we find one state
of the odd s-parity and the other of the even s-parity. At  $B=0$  the electron in the ground-state (first-excite-state) doublet
occupies predominantly a bonding (antibonding) orbital.
With the solid (dashed) lines we plotted the even (odd) s-parity energy levels.
Black lines show the results without SO coupling.
The blue lines correspond to the case of SO coupling without the linear Rashba term ($H_{SIA}^{lin}$), i.e. for $F_z=0$.
The red curves in Fig. \ref{bab1}(a) correspond to $F_z=188.8$ kV/cm, for which the linear two-dimensional Rashba constant
is as large as the linear two-dimensional Dresselhaus one. Beyond increased width of the avoided crossing
no qualitative difference in the results obtained for these two values of $F_z$ is found. Therefore, below we assume $F_z=0$ unless stated otherwise.

\begin{figure}[ht!]
 \hbox{\rotatebox{0}{\epsfxsize=60mm
                \epsfbox[36 197 575 661] {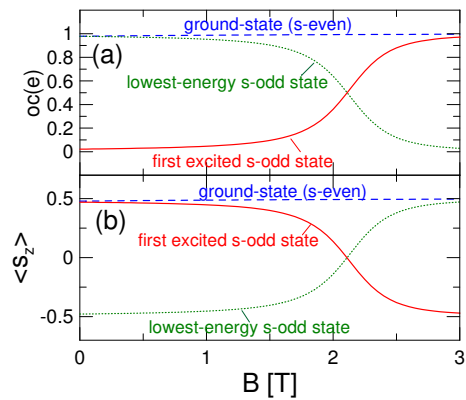}
                \hfill}}
\caption{Contribution of even-parity orbitals (a) and average value of the $z$ component of the spin (b) (in ${\hbar}$ units) in the lowest energy s-even and s-odd parity eigenstates.}
\label{1e4}
\end{figure}

\begin{figure}[ht!]
 \hbox{\rotatebox{0}{\epsfxsize=60mm
                \epsfbox[87 12 519 826] {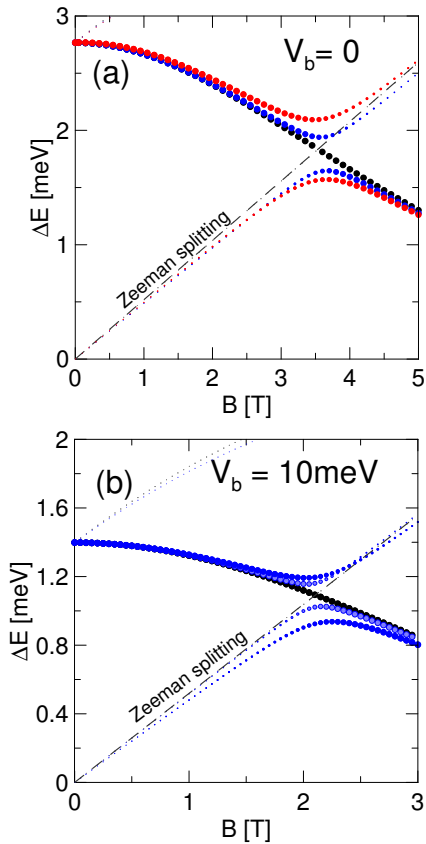}
                \hfill}}
\caption{
The dots show the low-energy absorption from the ground-state at $F_x=0$
for the single dot (a) and for the coupled dots (b) (for the energy spectra see Fig. \ref{bab1}).
The area of the dots is proportional to the absorption probability.
The black dots show the results without SO coupling. The full blue dots correspond to SO coupling with $F_z=0$.
 The open blue circles in (b) correspond
to height of the dot $d$ increased from 5.42 to 7.67 nm which amounts in a two-fold reduction of the 2D Dresselhaus constant.
The red dots in (a) correspond to a strong linear Rashba coupling present $F_z=188$ kV/cm.
The dashed grey line indicates the Zeeman splitting $g\mu_b B$.
}
\label{abs1e}
\end{figure}

\begin{figure}[ht!]
 \hbox{\rotatebox{0}{\epsfxsize=50mm
                \epsfbox[80 15 540 834] {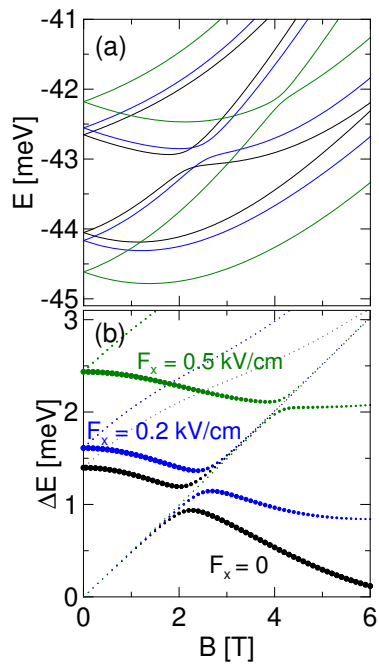}
                \hfill}}
\caption{Single-electron energy spectrum (a) and optical absorption spectrum (b) in function of the magnetic field for $F_x=0$ (black color), $F_x=0.2$ kV/cm (blue) and $F_x=0.5$ kV/cm (green) for coupled quantum dots.}
\label{1e5}
\end{figure}

For illustration of the double-dot wave functions we assumed a presence of a residual magnetic field $B=10\mu$T which lifts the doublet degeneracy
and we chose the states of the ground and excited doublets that correspond to $\langle s_z\rangle >0$.
With the blue lines in Fig. \ref{bab} we plotted the spinor components of the even s-parity ground state
which is bonding in its spin-up component with or without SO coupling. Its antibonding spin-down component appears when the SO
coupling is introduced  [Fig. \ref{bab}(c)].
The red lines in Fig. \ref{bab} correspond to the odd s-parity state of the excited doublet
which is antibonding in the spin-up component. The SO coupling adds to this state a bonding spin-down component.

In Fig. \ref{bab1} one observes an avoided crossing of
two excited energy levels of the odd s-parity stemming of both the ground and the exited  doublets.
Without the SO coupling the energy level that goes up in the energy with growing magnetic field
corresponds to the spin-down bonding orbital, and the one that goes down -- to the spin-up antibonding orbital.
The avoided crossing opened by the SO interaction is accompanied by spin and spatial parity mixing.

%The SO coupling leads to an appearance of an antibinding admixture in the single-electron ground state
%and an admixture of a bonding state in the first excited state.
For the single electron in ideally symmetric pair of dots ($F_x=0$) there is a direct correspondence between the SO-coupling-induced mixing of both the spin states and the occupation of
molecular orbitals of opposite spatial parity.
The occupation of the even parity orbitals [$\mathrm{oc}(e)$] is calculated as the norm of this component of the spinor that corresponds to the even parity state. Then
the average value of the $z$-component of the electron spin is
$\langle s_z\rangle=\hbar\left(\mathrm{oc}(e)-\frac{1}{2}\right)$ for the even s-parity and $\langle s_z\rangle=\hbar\left(\frac{1}{2}-\mathrm{oc}(e)\right)$ for
the odd s-parity states.
Occupation of the even parity orbitals and $\langle s_z\rangle$ is for the double dot displayed in Fig. \ref{1e4} as function of the magnetic field.
The ground state at higher field becomes a pure bonding spin-up orbital.
 We notice that the values corresponding to the two odd s-parity energy levels interchange near 2 T which is related
 to the energy level anticrossing presented in Fig. \ref{bab1}(b). At the center of the avoided crossing these two energy levels
 correspond to $\langle s_z\rangle=0$ and bonding and antibonding orbitals are equally occupied.

%In the absence of spin-orbit coupling one obtains a crossing of spin-down binding energy level with spin-up antibinding energy level near 3.5 T for $V_b=0$ [Fig. \ref{bab1}(a)] and near 2 T for $V_b=10$ meV [Fig. \ref{bab1}(b)]. The crossing energy levels correspond to the same s-parity and the spin-orbit
%coupling opens an avoided crossing between them.
%This avoided crossing is observed in the optical absorption spectrum presented in Fig. \ref{abs1e}.
%The spinor components in the center of the avoided crossing for $V_b=10$ meV are plotted
%in Figs. \ref{1e5a}(a,b).

%For $F=0$ optical transitions occur between states of opposite s-parities.

The discussed anticrossing of the odd s-parity energy levels leaves a clear signature on the optical
absorption spectrum. The energy and probability of excitation from the ground-state
are displayed in Fig. \ref{abs1e}.
The ground-state has the even s-parity hence the absorption is only allowed to the odd s-parity final state.
The ground-state is nearly spin-up polarized (Fig. \ref{1e4}) and since electron spin is left unchanged during an optical transition
 the absorption goes to the s-odd state with spin-up orientation.
When the avoided crossing is opened between the s-odd energy levels both of them possess a non-zero spin-up component
and the optical transitions to both of them from the ground-state are allowed.
Outside the avoided crossing the absorption spectra with or without SO coupling are similar.

%The SO coupling leaves a clear signature on the low-energy absorption spectrum of which is similar for a single elongated dot and
%for coupled quantum dots [Fig. \ref{abs1e}].  The energy splitting of the binding and antibinding energy levels decreases with increasing $V_b$.
%For that reason in the double dot the absorption energy range is reduced twice with respect to the
%case of a single elongated dot.
The energy range in which the SO-induced avoided crossing is observed in the absorption spectrum corresponds to far-infrared or microwave radiation
in which cyclotron resonance experiments are performed.\cite{microfal} One can increase the energy of the avoided crossing
twice by applying an electric field of 0.5 kV/cm - see Fig. \ref{1e5}(b). In the presence of the electric field
the electron in the ground-state is pushed to the left dot by $F_x>0$ while the final state in the absorption
process is mainly localized in the right dot [see Fig. \ref{bab}(b)].
The opposite shifts of the electron wave function in the initial and final states
are translated by the electric field into an  increased transition energy [see Fig. \ref{1e5}(a) for the energy splitting].
The obtained energy increase is accompanied by reduction of the SO-induced avoided crossing.

Fig. \ref{1e5}(b) shows also that for non-zero $F$ the absorption probabilities
vanish at higher $B$. The separation of the initial and final states [Fig. \ref{bab}(b)] by the electric field is enhanced
 when the magnetic field is applied, since the latter increases the localization of wave functions near the centers of the dots
  lifting the interdot tunnel coupling. In consequence
- the ground-state becomes totally localized in one dot
and the final state of the transition in the other. Vanishing overlap between the
initial and final state wave function implies vanishing transition probability as calculated by formula (12).

%In Fig. \ref{abs1e} and Fig. \ref{bab1}(a) with the red color we plotted the results for a strong inversional asymmetry of the structure,
%for which the Rashba linear constant is equal to the linear Dresselhaus constant.
%For the applied material parameters the results with the strong
%Rashba coupling are not qualitatively different from the case when the linear Rashba constant is zero.
%In the rest of the results for simplicity we assume that no inversional asymmetry of the structure is present.
%In fact the spin-orbit coupling is dominated by the linear Dresselhaus coupling.
%We find that the effects of all the other terms in the SO coupling are very small.

\begin{figure}[ht!]
 \hbox{\rotatebox{0}{\epsfxsize=60mm
                \epsfbox[60 15 545 827] {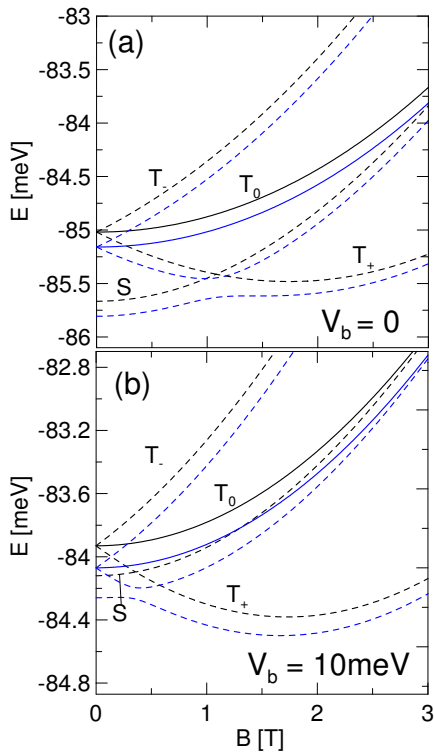}
                \hfill}}
\caption{Two-electron energy spectrum for a single elongated dot $V_b=0$ (a) and for a couple of dots separated by $V_b=10$ meV barrier (b).
Black (blue) lines show the results without (with) SO coupling. For the results without SO coupling we added labels $S$ for the singlet and $T$ for the triplets (subscript denotes the sign of the $z$-component of the total spin). Dashed (solid) curves correspond to odd (even) s-parity. Results were obtained for $F_x=0$.}
\label{2es}
\end{figure}

\begin{figure}[ht!]
 \hbox{\rotatebox{0}{\epsfxsize=80mm
                \epsfbox[28 133 557 696] {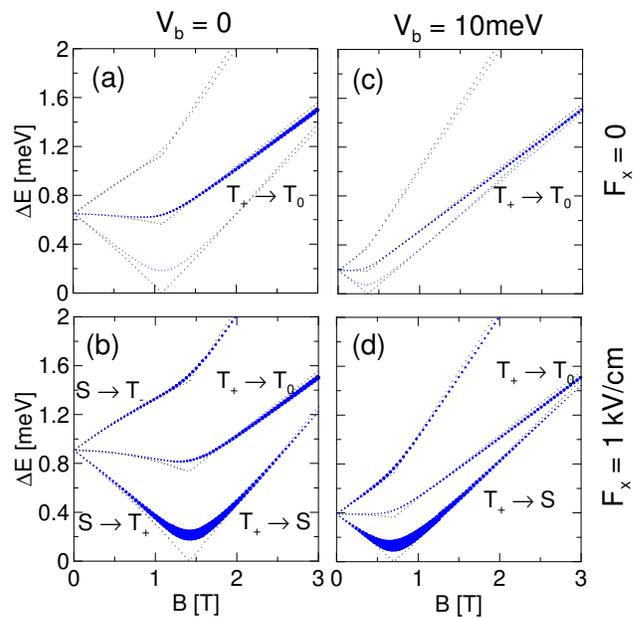}
                \hfill}}
\caption{Two-electron ground-state absorption spectrum as a function of the magnetic field for SO coupled single dot (a,b)
and double dot (c,d). Panels (a,c) correspond to $F_x=0$ and (b,d) to $F_x=1$ kV/cm. The area of the dots is proportional to the absorption probability.
Transitions are denoted by labels of two-electron spin eigenstates which are found without SO coupling. Without SO coupling
all the transitions presented in this figure are forbidden.
}
\label{2et}
\end{figure}

\begin{figure}[ht!]
 \hbox{\rotatebox{0}{\epsfxsize=40mm
                \epsfbox[28 133 578 717] {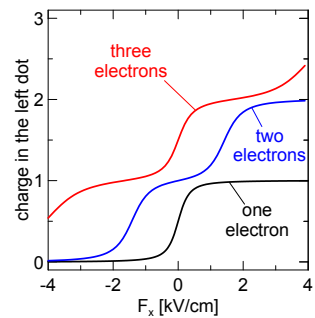}
                \hfill}}
\caption{Electron charge localized in the left dot as a function of the electric field for the double dot at $B=0$. Results
with and without SO coupling are not distinguishable.}
\label{lb0}
\end{figure}

\begin{figure*}[ht!]
 \hbox{\rotatebox{0}{\epsfxsize=140mm
                \epsfbox[10 328 578 526] {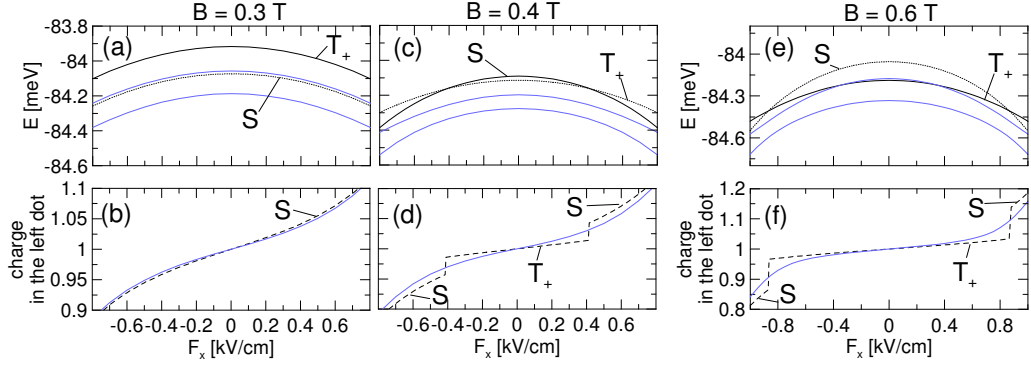}
                \hfill}}
\caption{Two-electron energy spectrum of the double dot is shown in (a,c,e) as
a function of the electric field.
Plots (b,d,f) indicate the charge localized in the left dot. Black (blue) lines show the results without (with) SO coupling.
The labels $S$ and $T_+$ correspond to eigenstates without SO coupling.}
\label{2e7}
\end{figure*}

\begin{figure}[ht!]
 \hbox{\rotatebox{0}{\epsfxsize=55mm
                \epsfbox[31 139 585 700] {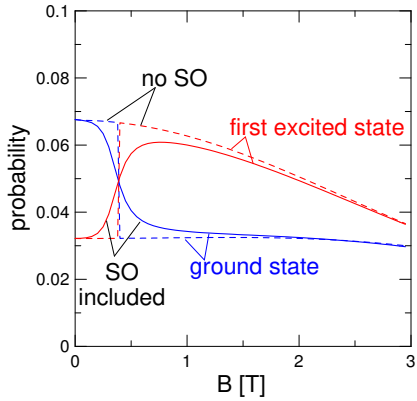}
                \hfill}}
\caption{Probability that both the electrons occupy the same dot with and without SO coupling in the ground state and first excited state.}
\label{2e4}
\end{figure}

\begin{figure}[ht!]
 \hbox{\rotatebox{0}{\epsfxsize=60mm
                \epsfbox[21 150 577 700] {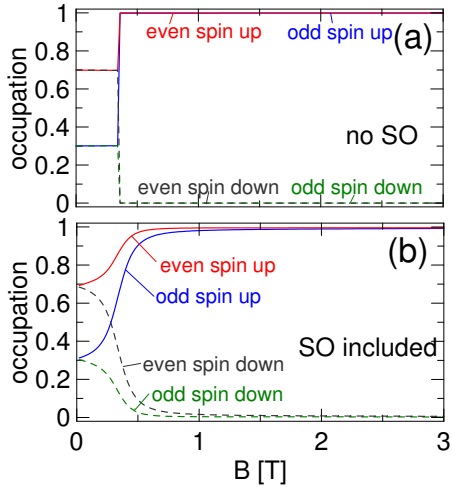}
                \hfill}}
\caption{Occupation of the single-electron orbitals of definite spatial parity and spin for two-electron ground state
with (b) and without (a) SO coupling for the electron pair in the double dot. }
\label{cc}
\end{figure}

\begin{figure}[ht!]
 \hbox{\rotatebox{0}{\epsfxsize=60mm
                \epsfbox[21 150 577 700] {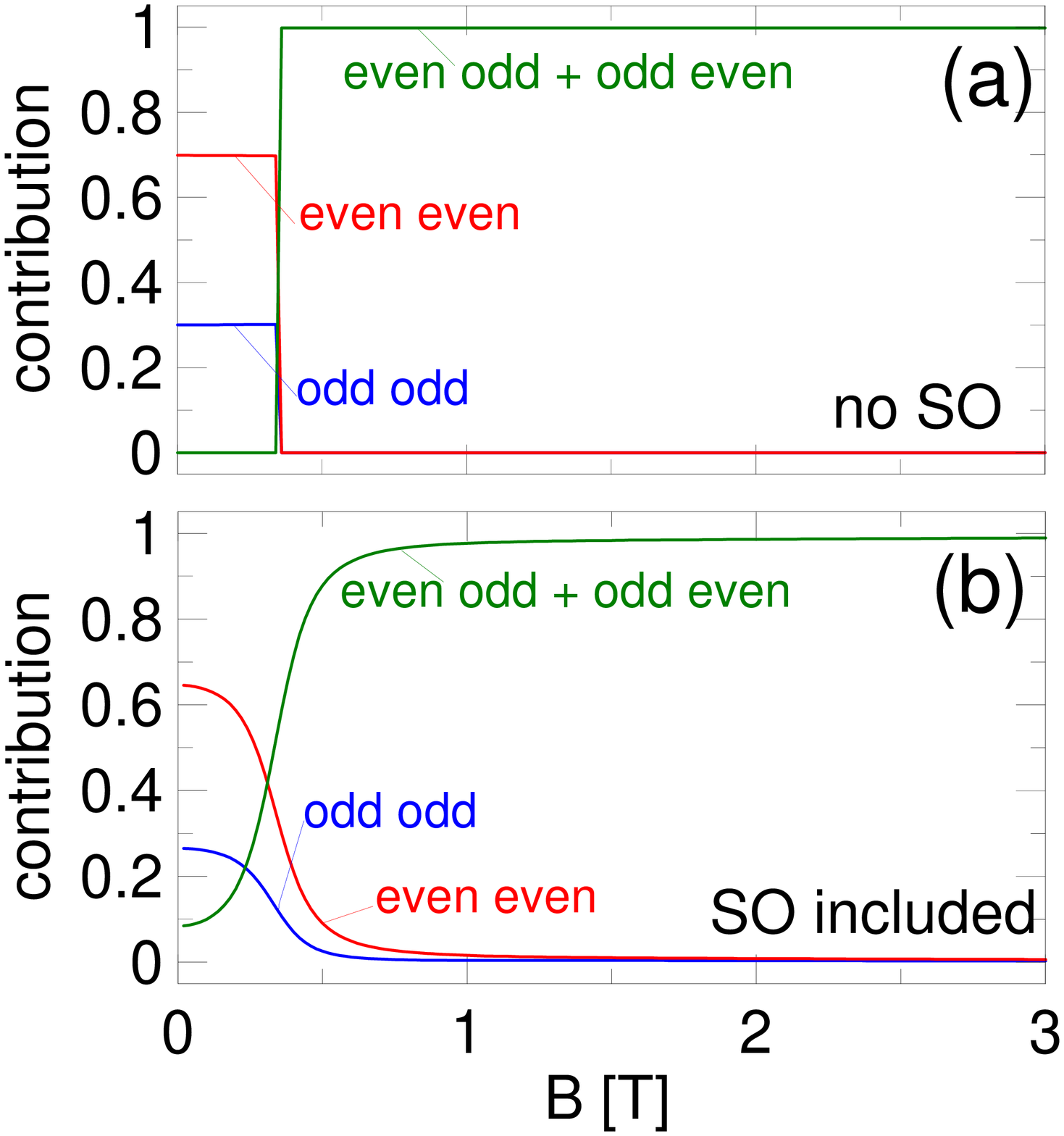}
                \hfill}}
\caption{Contributions of the two-electron orbitals to the ground state
with (b) and without (a) SO coupling for the electron pair in the double dot. }
\label{c2c}
\end{figure}

\subsection{Electron pair}

In the absence of the magnetic field and without SO coupling the first excited state of the electron pair is spin triplet.
For $B=0$ we find that the first excited state is threefold degenerate also with SO coupling present.
This applies to both the single elongated dot [Fig. \ref{2es}(a)] and the double dot [Fig. \ref{2es}(b)].
Without SO coupling the magnetic field induces
a singlet-triplet ground-state transition near 1 T for the single dot and near 0.4 T for the double dot.
The crossing singlet and triplet energy levels have the same odd s-parity and an avoided crossing is opened between them when SO coupling is introduced.
The calculated width of the avoided crossing is 0.18 and 0.07 meV for the single and double dot, respectively
 which is within the order of the ones found in experiments: 0.25 meV and 0.2 meV for the nanowire quantum dot\cite{stxp}
 and for the double dots\cite{stxp2}. 

For a symmetric system ($F_x=0$) the optical transition from the ground-state can only go to the even s-parity  eigenstate.
In the absence of the spin-orbit coupling in the considered energy range only the triplet with zero $z$-component of the spin ($T_0$)
has the required spatial parity to absorb photons. However, this absorption is excluded anyway
on both sides of the singlet-triplet ground-state transition. For $B$ below this transition the matrix element (12)
vanishes due to opposite symmetry of the spatial initial and final wave functions with respect to the electron interchange.
For $B$ above the singlet-triplet transition the ground-state (triplet with $s_z=\hbar$ denoted as $T_+$) and $T_0$ states have the same symmetry with respect to the
electron interchange, but the $z$-components of the spin are different.
Optical transitions between the states corresponding to energy levels presented in Fig. \ref{2es} are only allowed by the SO coupling.
The calculated absorption spectrum is shown in Fig. \ref{2et}. For $F_x=0$ [Fig. \ref{2et}(a,c)] the absorption probability
grows with the magnetic field after the singlet-triplet ground state avoided crossing.
Then, the transition corresponds to $T_+\rightarrow T_0$ excitation in terms of states without SO coupling.
When the electric field $F_x$ is switched on [Fig. \ref{2et}(b,d)] the  parity selection rules no longer apply and we notice appearance of also $S\leftrightarrow T_+$ and $S\rightarrow T_-$ transitions.
The probabilities for the discussed transitions -- which are all forbidden in the absence
of SO coupling -- remain very small (less than 0.5\%) as compared to the ones found for the single and three electrons.

For two electrons the role of the electric field for the low-energy optical absorption
is different from the single-electron case. For $N=1$ the electric field distinctly shifts the energy of the
absorption lines (Fig. 6). For $N=2$ the energy shift is very weak, only the transition probabilities are affected. For the single electron
the energy shifts resulted from spatial electron-charge displacements of the initial and final states induced by the electric field. For two electrons
these shifts are hampered (see Fig. 1) since the charge shift implies appearance of a double occupation of one of the dots.
Fig. \ref{lb0} shows the charge localized in the left dot in function of the electric field. For $N=1$ (and $N=3$) the dependence
of the charge on $F_x$ is the strongest at zero electric field, while for $N=2$ we find a plateau centered at $F_x=0$.

For $B=0$ we did not find
any SO coupling influence on the charge distribution as a function of the in-plane electric field. Nevertheless, such an effect is observed in the presence of the external magnetic field - see Fig. \ref{2e7}.
For $B=0.4$ T the ground-state without SO coupling corresponds already to the spin triplet, in which -- due to the Pauli exclusion -- localization of both electrons
in the same dot requires occupation of an excited single-dot energy level.
The charge of the two-electron system for the triplet ground-state is even more resistant to shifts by the electric field
than for the singlet state [compare Fig. \ref{2e7}(b) and (d)]. For  $B=0.4$ T the ground-state
becomes singlet again near 0.4 kV/cm. The electrons in the singlet state occupy more easily \cite{szafran} the dot made deeper by the electric field which restores the singlet ground-state when $F_x$ is applied. We notice [see the dashed line in Fig. \ref{2e7}(d)] a jump in the occupation of the left dot at the singlet-triplet transition. For $B=0.6$ T a similar effect
is observed only at higher $F_x$ [the dashed line in Fig. \ref{2e7}(f)]. The SO coupling mixes the singlet and triplet states and we notice that the electron charge in the left dot [blue lines in Fig. \ref{2e7}(b,d,f)]
becomes a smooth function of $F_x$. As a general rule, when the ground-state without SO coupling is singlet (triplet) - the SO coupling reduces (enhances) the occupation of the deeper dot.

At the singlet-triplet transition the SO coupling influences also the probability of finding both the electrons
in the same dot (Fig. \ref{2e4}). Without SO coupling the ground-state probability exhibits a rapid drop at the singlet-triplet transition near 0.4 T.
The spin-orbit coupling influences the double occupation probability only for non-zero $B$.

In order to quantify the occupation of the single-electron even and odd parity orbitals we first
project the two-electron eigenstates of operator (11) into the basis composed of single-electron eigenfunctions obtained without SO coupling
(denoted as $\psi'$ in the following). For a state $\nu$ we consider the projection in form
\begin{eqnarray}
d_{kl}^\nu&=&\frac{1}{2}\sum_{i,j>i}  C^\nu_{ij} \langle  \psi_i(1)\psi_j(2) - \psi_i(2)\psi_j(1) | \nonumber \\ &&|\psi^{'}_k(1)\psi^{'}_l(2) - \psi^{'}_k(2)\psi^{'}_l(1)\rangle.
\end{eqnarray}
An eigenfunction $\psi'_k$ has a definite spatial parity and $z$-component of the spin associated with a spinor $\chi_k$ which is
the $s_z$ eigenfunction of eigenvalue $\hbar/2$ or $-\hbar/2$  ($\chi_k=|\uparrow\rangle$ or $\chi_k=|\downarrow\rangle$).
Hence, the occupation of the spin-up even-parity single-electron wave functions can be calculated as
\begin{equation}
\mathrm{oc}({e\uparrow})=\sum_{k,l>k} |d_{kl}|^2 \left[ \delta_p(k,+)\delta_s(k,\uparrow) + \delta_p(l,+)\delta_s(l,\uparrow)\right],
\end{equation}
where
\begin{equation} \delta_p(k,\pm)=\frac{1\pm\int\left(\psi'_k(r)\right)^*\psi'_k(-r)dr}{2}\end{equation}
and
\begin{equation} \delta_s(k,\uparrow)=\langle\chi_k|\uparrow\rangle .\end{equation}
The occupation of the spin-up odd-parity single-electron states is determined by the formula
\begin{equation}
\mathrm{oc}({o\uparrow})=\sum_{k,l>k} |d_{kl}|^2 \left[ \delta_p(k,-)\delta_s(k,\uparrow) + \delta_p(l,-)\delta_s(l,\uparrow)\right],
\end{equation}
with an obvious generalization for the spin-down components.
The results are displayed in Fig. \ref{cc}.
Without SO coupling  i) below 0.4 T the ground-state is even parity singlet -
the electrons occupy mostly the even parity states ii)  above 0.4 T the
ground-state is odd parity triplet - the spin-down contributions are removed, one of the
electrons occupy an even parity and the other an odd parity orbital.
The jump of the occupations near 0.4 T that is observed in the results without SO coupling is replaced by a smooth transition
when SO coupling is applied. The values obtained for orbital occupations in both large and zero $B$ limits are similar.

Non-conservation of the spatial parity in the presence of SO coupling for the two-electron states becomes evident
when one considers contributions of the two-electron basis elements. The contributions of the elements in which both
electrons occupy orbitals of the same spatial parity are calculated as
\begin{equation}
c_{ee}=\sum_{k,l>k} |d_{kl}|^2 \delta_p(k,+)\delta_p(l,+),
\end{equation}
for the even parity orbitals and
\begin{equation}
c_{oo}=\sum_{k,l>k} |d_{kl}|^2 \delta_p(k,-)\delta_p(l,-),
\end{equation}
for the odd parity orbitals.
Contribution of the two-electron basis elements in which the electrons occupy opposite parities is
\begin{equation}
c_{oe+eo}=\sum_{k,l>k} |d_{kl}|^2 \left[ \delta_p(k,-)\delta_p(l,+) + \delta_p(k,+)\delta_p(l,-)\right].
\end{equation}
The results are displayed in Fig. \ref{c2c}.
Without SO coupling for $B<0.4$ T  the contribution of the basis
elements in which the electrons occupy opposite parity eigenstates is zero. In the triplet ground state for $B>0.4$ T
the electrons are bound to occupy orbitals of opposite parities.
When the SO is present for $B=0$ there is a nearly  10\% contribution of basis elements in which the electrons
occupy orbitals of opposite parities.
The $c_{oe+eo}$ grows with the magnetic field, but it stays below 100\% in the studied
range of $B$. This result and the ones presented above indicate that for two electrons the SO coupling has a noticeable influence
on the ground-state properties in contrast to the single electron case.

\subsection{Three electrons}

\begin{figure}[ht!]
 \hbox{\rotatebox{0}{\epsfxsize=55mm
                \epsfbox[28  150 565 682] {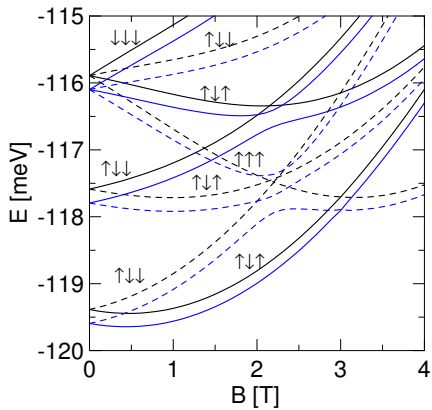}
                \hfill}}
\caption{Three-electron energy spectrum for the single elongated dot. Black (blue) lines show the results
without (with) SO coupling. Solid (dashed) lines correspond to even (odd) s-parity states. The arrows in the plot
indicate the $z$ component of the spin without SO coupling.}
\label{3e1}
\end{figure}

\begin{figure}[ht!]
 \hbox{\rotatebox{0}{\epsfxsize=55mm
                \epsfbox[24  42 565 821] {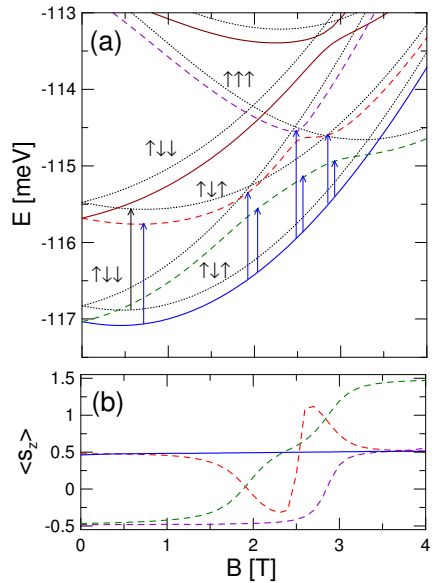}
                \hfill}}
\caption{Three-electron energy spectrum (a) for the double dot. Black (color) lines show the results
without (with) SO coupling. Solid (dashed) lines correspond to even (odd) s-parity. The short arrows in the plot
indicate the $z$ component of the spin without SO coupling and the longer ones show the allowed optical transitions from the ground-state.
(b) $z$ component of the spin for the lowest even s-parity and three lowest odd s-parity energy levels. Type and color of curves for these states is adopted of panel (a).
}
\label{3e3}
\end{figure}

For $N=3$ in the absence of SO coupling the magnetic field leads to the ground-state spin-polarization transition near 3 T in both the single
(Fig. \ref{3e1}) and double [Fig. \ref{3e3}(a)] dots.
For symmetric dots this transition is associated with energy level crossing even
when SO coupling is introduced since the ground-states on both sides of the transition correspond
to opposite s-parities.
The in-plane electric field opens an avoided crossing at the ground-state spin polarization transition [see Fig. \ref{3e5}(a)].

For three electrons in a single dot without SO coupling one observes (Fig. \ref{3e1}) crossings of three s-odd energy levels near 2 T. For the double dot [Fig. \ref{3e3}(a)]
the crossings appear in more separated magnetic fields. The three crossing levels have different $z$ projections of the spin.
Similarly as for $N=1$ the SO coupling opens avoided crossing in the absorption spectrum, but for $N=3$ three energy levels
participate in this avoided crossing instead of two.
These avoided crossings are accompanied by a smooth variation of the spin [Fig. \ref{3e3}(b)].

\begin{figure}[ht!]
 \hbox{\rotatebox{0}{\epsfxsize=85mm
                \epsfbox[28  250 570 570] {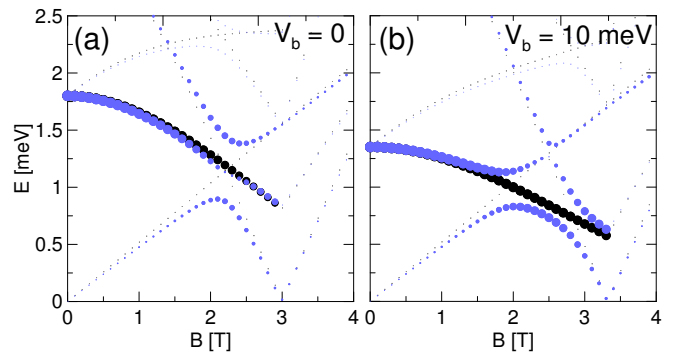}
                \hfill}}
\caption{Optical absorption spectrum for the three-electron system in the single dot (a) and in the double dot (b).
Black (blue) dots correspond to SO coupling absent (present). Area of the dot is proportional to the absorption probability.}
\label{3e2}
\end{figure}

\begin{figure}[ht!]
 \hbox{\rotatebox{0}{\epsfxsize=55mm
                \epsfbox[86  17 536 826] {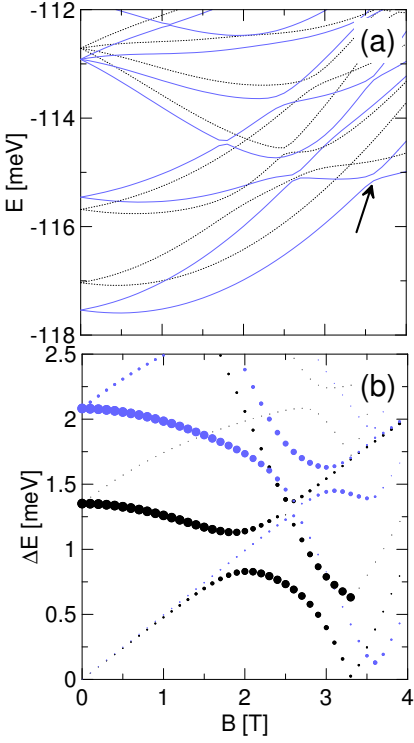}
                \hfill}}
\caption{(a) Three electron energy spectrum for SO-coupled double dot at $F_x=0$ (black dotted lines)
and for $F_x=0.5$ kV/cm (blue solid curves). The arrow indicates the ground-state
avoided crossing which is opened in presence of non-zero $F_x$.
(b)
Optical absorption spectrum for the SO-coupled double dot. Black (blue) dots correspond to $F_x=0$ ($F_x=0.5$ kV/cm). }
\label{3e5}
\end{figure}

For the spin unpolarized ground-state ($B<3$ T) the lowest-energy optical transition goes from the ground-state
to the odd s-parity states with $\langle s_z\rangle\simeq \hbar/2$.
Without SO coupling and in terms of occupation of single-electron orbitals we observe (Fig. \ref{3e2}) a transition of one of the electrons occupying a bonding
orbital to an occupied antibonding orbital. One finds a single bright line similar to the one found for $N=1$.
For $B>3$ T the principle line in the
ground-state absorption spectrum disappears due to the ground-state spin polarization.

The in-plane electric field increases the energy splitting between the ground-state and the first excited state leading to an increase
of the energy absorbed at the optical transition [Fig. \ref{3e5}(b)]. The form of the avoided crossing is
not affected by the field -- like in the single electron case.

\begin{figure}[ht!]
 \hbox{\rotatebox{0}{\epsfxsize=85mm
                \epsfbox[20  252 586 600] {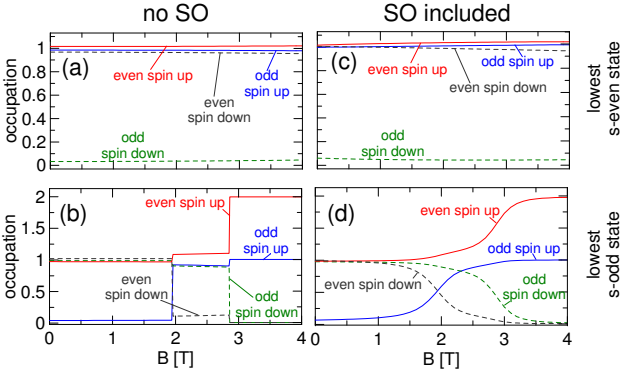}
                \hfill}}
\caption{Occupation of the single-electron orbitals of definite spin orientation and spatial parity
without (a,b) and with (c,d) SO coupling for the lowest energy s-even  (a,c) and s-odd (b,d) state
for three electrons in the double dot.}
\label{3e4}
\end{figure}

\begin{figure}[ht!]
 \hbox{\rotatebox{0}{\epsfxsize=85mm
                \epsfbox[20  252 586 600] {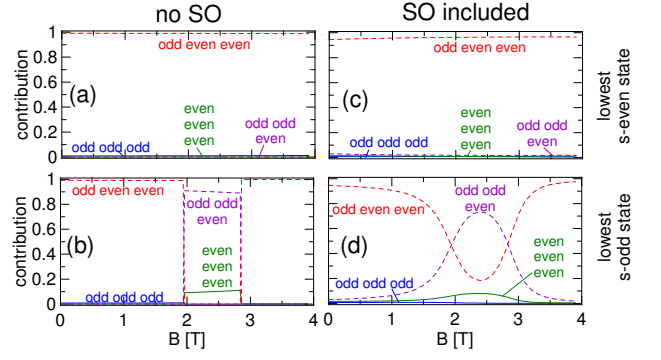}
                \hfill}}
\caption{Contributions of single-electron orbitals of a given symmetry to the 
lowest energy three electron s-even (a,c) and s-odd (b,d) states,
with (c,d) and without (a,b) SO coupling.}
\label{3e4a}
\end{figure}

 For the lowest-energy even s-parity state  both occupation of single-electron spin-orbitals [Fig. \ref{3e4}(a,c)] and
contribution of three-electron basis elements of definite spatial parity [Fig. \ref{3e4a}(a,c)]  are only
weakly affected by both the magnetic field
and the spin-orbit coupling. The dependence of the studied quantities on the magnetic field is more spectacular
for the lowest energy s-odd state [Figs. \ref{3e4}(b,d) and \ref{3e4a}(b,d)].
Without  SO coupling the lowest-energy s-odd level corresponds to even parity only between 1.9 and 2.8 T,
hence the vanishing contribution of the even-parity three-electron basis elements outside this $B$ interval.
In the presence of SO coupling the contribution of the even-parity basis elements extends over the entire studied range of the magnetic field.

\section{Summary and Conclusions}
We have presented a systematic exact diagonalization study of one, two and three-electron spin-orbit coupled systems in double quantum dots.
We discussed the mixing of the bonding and antibonding electron orbitals by the SO coupling.
We investigated occupation of even and odd parity orbitals, the energy and optical absorption spectra in crossed electric and magnetic fields
as well as the electron distribution.

For one and three electrons confined in a pair of identical dots we found that the spin-orbit coupling only weakly
affects the ground-state properties. A strong mixing of bonding and antibonding orbitals due to the spin-orbit coupling was found in the lowest-energy
excited states.

In contrast to the odd electron numbers, for two electrons the spin-orbit interaction affects the properties
of the ground-state since the spin-polarization becomes a smooth transition instead of an abrupt
singlet-triplet transformation. On the contrary, the spin polarization of the three electron system in symmetric dots
is not affected by the spin-orbit coupling since the low- and high-spin ground-states correspond to opposite s-parities.
For three electrons the SO coupling makes the  spin-polarization continuous only when the
confinement potential contains an in-plane asymmetry, e.g. introduced by an electric field.

For odd electron numbers the spin-orbit-coupling-induced mixing of spatial parities of the first excited state opens
characteristic avoided crossings in the optical absorption spectrum. An in-plane electric field shifts the initial and final states of the optical transition
to opposite dots.
In consequence it distinctly increases the energy of the optical transition at an expense of a reduced width of the avoided crossing.

The low-energy optical absorption for two electrons is
only allowed by the SO coupling. For two electrons the in-plane electric field lifts the spatial parity selection rules but does not essentially perturb the energy of the optical transitions.

\acknowledgments{This work was supported by the ''Krakow Interdisciplinary
PhD-Project in Nanoscience and Advanced Nanostructures"
operated within the Foundation for Polish Science
MPD Programme co-financed by the EU European Regional
Development Fund.}

\end{document}